\documentclass[final,5p,twocolumn,times,sort&compress]{elsarticle}

\usepackage{epsfig}
\usepackage{amsmath}
\usepackage{slashed}

\newif\ifContLineOne
\newif\ifContLineTwo
\newif\ifContLineThree

\def\conC#1{\vbox{\ialign{##\crcr
  \ifContLineThree\hrulefill\else\vphantom{\hrulefill}\fi\crcr
  \noalign{\kern3.2pt\nointerlineskip}
  \ifContLineTwo\hrulefill\else\vphantom{\hrulefill}\fi\crcr
  \noalign{\kern3.2pt\nointerlineskip}
  \ifContLineOne\hrulefill\else\vphantom{\hrulefill}\fi\crcr
  \noalign{\nointerlineskip}
  $\hfil\textstyle{\vbox to 14pt{}#1}\hfil$\crcr}}}

\def\DrawLeg#1#2{
  \kern-.2pt              
  \dimen2 =#1             
  \advance\dimen2 by 2pt  
  \dimen3 = 10.6pt        
  \dimen4 =3.6pt          
  \advance\dimen3 by -\dimen2 
  \multiply\dimen4 by #2
  \advance\dimen3 by \dimen4
  \raise\dimen2 \hbox{\vrule height\dimen3 width .4pt} 
  \kern-.2pt}             

\def\begC#1#2{\setbox0 =\hbox{$\textstyle{#2}$}
  \dimen0=.5\wd0 \dimen1=\ht0
  \conC{\hskip\dimen0}
  \count255=#1
  \ifnum\count255 =1 \ContLineOnetrue\else
  \ifnum\count255 =2 \ContLineTwotrue\else
  \ifnum\count255 =3 \ContLineThreetrue\fi\fi\fi
  \DrawLeg{\dimen1}{\count255}
  \conC{\hskip\dimen0}
  \kern-\dimen0\kern-\dimen0 \box0}

\def\endC#1#2{\setbox0 =\hbox{$\textstyle{#2}$}
  \dimen0=.5\wd0 \dimen1=\ht0
  \conC{\hskip\dimen0}
  \count255=#1
  \ifnum\count255 =1 \ContLineOnefalse\else
  \ifnum\count255 =2 \ContLineTwofalse\else
  \ifnum\count255 =3 \ContLineThreefalse\fi\fi\fi
  \DrawLeg{\dimen1}{\count255}
  \conC{\hskip\dimen0}
  \kern-\dimen0\kern-\dimen0 \box0}

\def\preprintdate{IUHET 615, October 2016}

\def\al{\alpha}
\def\be{\beta}
\def\ga{\gamma}
\def\de{\delta}
\def\ep{\epsilon}

\def\et{\eta}
\def\th{\theta}

\def\la{\lambda}

\def\si{\sigma}

\def\ph{\phi}
\def\vp{\varphi}
\def\ch{\chi}
\def\ps{\psi}
\def\om{\omega}
\def\Ga{\Gamma}

\def\cL{{\cal L}}
\def\cM{{\cal M}}

\def\cR{{\cal R}}

\newcommand{\beq}{\begin{equation}}
\newcommand{\eeq}{\end{equation}}
\newcommand{\bea}{\begin{eqnarray}}
\newcommand{\eea}{\end{eqnarray}}
\newcommand{\bit}{\begin{itemize}}
\newcommand{\eit}{\end{itemize}}
\newcommand{\rf}[1]{(\ref{#1})}

\def\fr#1#2{{{#1} \over {#2}}}
\def\half{{\textstyle{1\over 2}}}
\def\quar{{\textstyle{1\over 4}}}

\def\lsim{\mathrel{\rlap{\lower4pt\hbox{\hskip1pt$\sim$}}
    \raise1pt\hbox{$<$}}}
\def\gsim{\mathrel{\rlap{\lower4pt\hbox{\hskip1pt$\sim$}}
    \raise1pt\hbox{$>$}}}
\def\sqr#1#2{{\vcenter{\vbox{\hrule height.#2pt
         \hbox{\vrule width.#2pt height#1pt \kern#1pt
         \vrule width.#2pt}
         \hrule height.#2pt}}}}

\def\prt{\partial}
\def\lrprt{\hskip-4pt\stackrel{\leftrightarrow}{\prt}}

\def\Im{\hbox{Im}\,}
\def\tr{{\rm tr}~}
\def\etal{{\it et al.}}

\def\ol#1{\overline{#1}}

\def\psb{\ol{\ps}}
\def\wt#1{\widetilde{#1}}

\def\qq#1{q_{#1}}
\def\X{X}
\def\P{p}
\def\Q{q}
\def\ring#1{{\mathaccent'27 #1}}

\begin{document}

\begin{frontmatter}

\title{Lorentz violation and deep inelastic scattering}

\author{V.\ Alan Kosteleck\'y$^1$, E.\ Lunghi$^1$, and A.R.\ Vieira$^{2,3}$}

\address{$^1$Physics Department, Indiana University, 
Bloomington, IN 47405, USA\\
$^2$ Indiana University Center for Spacetime Symmetries, 
Bloomington, IN 47405, USA\\
$^3$ Departamento de F\'isica -- ICEx, 
Universidade Federal de Minas Gerais, 
Belo Horizonte, MG 30.161-970, Brasil}

\address{}
\address{\rm 
\preprintdate;
published as Phys.\ Lett.\ B {\bf 769}, 272 (2017)
}

\begin{abstract}
The effects of quark-sector Lorentz violation 
on deep inelastic electron-proton scattering are studied. 
We show that existing data can be used to establish first constraints
on numerous coefficients for Lorentz violation in the quark sector
at an estimated sensitivity of parts in a million.
\end{abstract}

\end{frontmatter}

\section{Introduction}
\label{sec:introduction}

Deep inelastic scattering (DIS) provides key experimental evidence
for the existence of quarks 
and the validity of quantum chromodynamics (QCD).
In early experiments on electron-proton scattering,
the DIS cross section was discovered 
to vary only weakly with momentum transfer
\cite{dis},
and the invariance of the DIS form factors under scaling
\cite{bjorken}
revealed that nucleons contain partons
\cite{feynman}.
Subsequent DIS studies with electrons and neutrinos
have confirmed this picture and verified predictions of QCD,
and DIS continues to be an essential tool in searches for new physics
\cite{disproc}.

One proposal for new physics 
is tiny observable violations of Lorentz invariance,
which could emerge as a byproduct of the unification
of gravity and quantum physics in a Planck-scale theory
such as strings
\cite{ksp}.
Many sensitive searches for Lorentz violations
have been performed in recent years,
spanning most sectors of the Standard Model (SM)
as well as gravity
\cite{tables}.
However,
direct information about the Lorentz properties of quarks
is comparatively difficult to obtain.
In the present work,
we investigate the prospects of using DIS 
as a tool to search for Lorentz violation.
We obtain dominant Lorentz-violating corrections 
to the DIS cross section for electron-proton scattering
and use data from the 
Hadronen-Elektronen Ring Anlage (HERA)
\cite{h1zeus}
to estimate attainable sensitivities
in a dedicated analysis searching for Lorentz violation.

Our treatment is based on techniques from effective field theory,
which is appropriate for investigating suppressed signals
from an experimentally inaccessible energy scale
\cite{sw}. 
The realistic effective field theory 
for general Lorentz violation is
known as the Standard-Model Extension (SME)
\cite{ck,akgrav}.
Each Lorentz-violating term in the SME action
is a coordinate-independent contraction
of a coefficient for Lorentz violation
with a Lorentz-violating operator,
which can be specified in part by its mass dimension $d$.
Adding all Lorentz-violating terms  
to the action for General Relativity coupled to the Standard Model
produces the SME action.
Restricting attention to terms with $d\leq 4$ in Minkowski spacetime
yields an action that is power-counting renormalizable,
called the minimal SME.
In realistic effective field theory,
violation of CPT symmetry implies Lorentz violation
\cite{ck,owg},
so the SME also characterizes general effects of CPT violation. 
For reviews of the SME see,
for example,
Refs.\ \cite{tables,reviews}.
Here,
we focus attention on the quark sector of the minimal SME,
seeking to identify its predictions for DIS.

One way to access quark-sector SME coefficients 
is to take advantage of the interferometric nature 
of neutral-meson propagation
\cite{ak98}.
Studies of kaon oscillations provide sensitivity to 
certain SME coefficients for CPT violation
involving the $d$ and $s$ quarks
\cite{kaons},
while oscillations of $D$, $B_d$, and $B_s$ mesons
have been used to constrain CPT violation involving 
also the $u$, $c$, and $b$ quarks
\cite{DBmesons}.
The $t$ quark decays before hadronizing
and so its Lorentz properties cannot be investigated 
via interferometric methods,
but observations of the production and decay of $t$-$\ol t$ pairs
have been used to constrain SME coefficients 
for CPT-even Lorentz violation
in the $t$-quark sector 
\cite{tquark},
and coefficients for CPT-odd Lorentz violation
are accessible to studies of single-$t$ production and decay
\cite{bkl16}.
A few ultrarelativistic constraints on quark-sector coefficients
have also been obtained from high-energy cosmic rays 
\cite{astro}.

The primary goal of the present work 
is to show explicitly that electron-proton DIS can access 
spin-independent and CPT-even Lorentz violation
involving $u$ and $d$ quarks,
about which direct information is lacking in the literature.
Indirect clues can be gleaned from other experiments 
involving nuclei or hadrons
\cite{tables},
but extracting direct quark-sector bounds from these
requires disentangling possible Lorentz-violating effects
from quarks, gluons, and sea constituents.
Theoretical techniques such as Lorentz-violating chiral perturbation theory,
which to date have been used to interrelate various hadron SME coefficients
\cite{lvchpt},
could in principle also shed light on this issue.
In any event,
electron-proton DIS is of particular interest in this context
because it is well measured,
can be calculated with sufficient accuracy,
and is comparatively simple in that only 
one of the two colliding species involves quarks.

\section{Setup}

\begin{figure}
\begin{center}
\includegraphics[width=0.7\hsize]{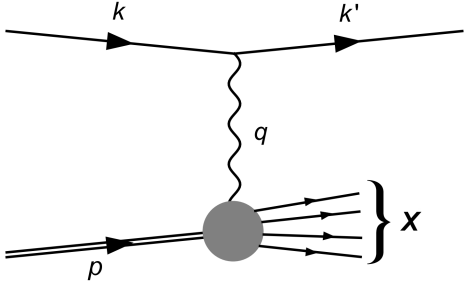}
\caption{
DIS of an electron (momentum $k^\mu$) off a proton (momentum $\P^\mu$).
\label{fig1}}
\end{center}
\end{figure}

Since DIS is a high-energy process
with momentum transfer greater than the proton mass,
asymptotic freedom allows a perturbative treatment.
At zeroth order in the strong coupling constant $g_s$,
the interactions among quarks are neglected 
and the photon exchanged between the electron and the proton 
interacts with partons carrying a fraction of the proton total momentum. 
We focus on calculating the tree-level impact 
of Lorentz violation on this process,
shown in Fig.\ \ref{fig1}.

For simplicity and definiteness,
we limit attention to dominant effects in unpolarized electron-proton DIS
and we neglect possible quark flavor-changing couplings.
It is reasonable also to neglect electron- and photon-sector Lorentz violation,
which are constrained well below levels attainable here
\cite{tables}.
The primary Lorentz-violating behavior can therefore be expected 
to arise from spin-independent CPT-even operators
in the quark sector of the minimal SME.
Including the two valence-quark flavors $f=u,d$ for the proton,
the relevant part of the Lagrange density 
for Lorentz-violating QCD 
augmented by electromagnetic couplings 
\cite{ck}
is then
\beq
\cL \supset 
\sum_{f=u,d}
(\et^{\mu\nu} + c^{\mu\nu}_f) 
\psb_f
(\half \ga_{\mu}i\lrprt_{\nu}
- \qq f \ga_{\mu}A_{\nu})
\ps_f.
\label{eq1}
\eeq
The dimensionless SME coefficients $c^{\mu\nu}_f$ 
control the magnitude of the Lorentz violation 
and can be taken as constant in an inertial frame
in the vicinity of the Earth
\cite{ck},
which insures energy and momentum remain conserved.
The coefficients modify the dispersion relation of the quarks,
which propagate along pseudo-Finsler geodesics
\cite{finsler}. 
In the high-energy limit $Q^2=-q^2 \gg M^2$, 
where $M$ is the mass of the proton,
we can disregard the strong interactions and view the quarks as interacting 
only with the photon through their charges $\qq f$.
Nonetheless,
the couplings in $\cL$ reveal that the DIS cross section
is affected by the quark-sector SME coefficients 
$c_f^{\mu\nu}$ 
in Lorentz-violating QCD.

The form of the Lagrange density $\cL$ is simplified
by the possibility of performing 
coordinate choices and field redefinitions 
that leave unaffected the observable physics
\cite{ck,akgrav,redefs}.
For example,
the contractions in the Maxwell term 
$-\quar F_{\mu\nu}F^{\mu\nu}$ 
could in principle be performed 
with an effective metric involving 
the photon-sector coefficient $(k_F)_\al{}^{\mu\al\nu}$ 
for Lorentz violation,
but a suitable choice of coordinates can always absorb
this into the coefficients $c_f^{\mu\nu}$ 
and the corresponding electron-sector coefficient $c_e^{\mu\nu}$.
If desired,
the general case can be recovered via the substitutions
$c_f^{\mu\nu} \to c_f^{\mu\nu} + \half (k_F)_\al{}^{\mu\al\nu}$
and $c_e^{\mu\nu} \to c_e^{\mu\nu} + \half (k_F)_\al{}^{\mu\al\nu}$.
In this context,
disregarding Lorentz violation in the electron and photon sectors
amounts to assuming that the combination
$c_e^{\mu\nu} + \half (k_F)_\al{}^{\mu\al\nu}$
is negligible based on existing precision tests of Lorentz invariance
\cite{tables}.
Field redefinitions also insure that spin-independent CPT-odd effects
are unobservable at this level,
and they imply that the coefficients
$c_f^{\mu\nu}$ 
can be taken as symmetric without loss of generality.
The trace $\et_{\mu\nu}c_f^{\mu\nu}$ produces no Lorentz-violating effects
and can be set to zero.
The Lorentz-violating physics of interest here 
is therefore controlled by nine independent components 
of $c_f^{\mu\nu}$ for each quark flavor $f$.

\begin{figure}
\begin{center}
\includegraphics[width=0.9\hsize]{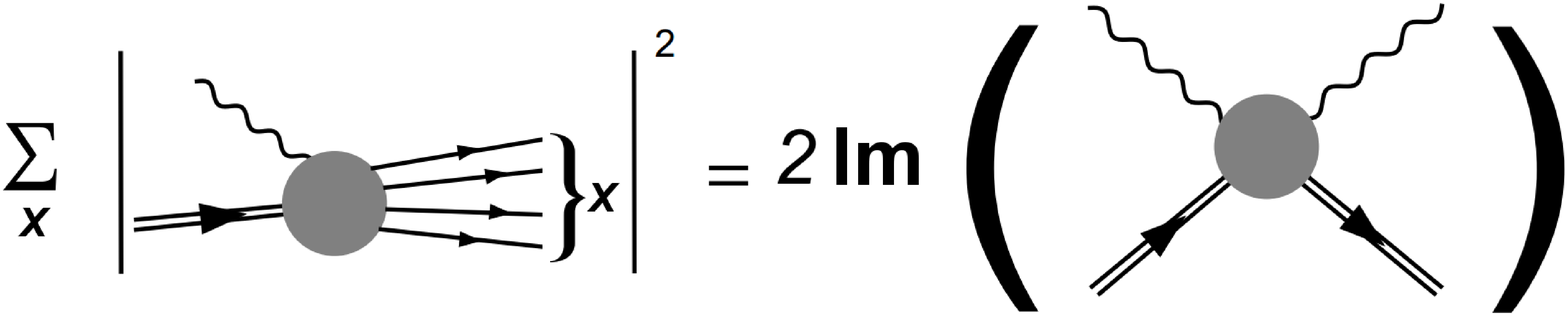}
\caption{
Optical theorem for the photon-proton interaction. 
\label{fig2}}
\end{center}
\end{figure}

To initiate the calculation,
we first consider the DIS amplitude of Fig.\ \ref{fig1}
for a single quark of flavor $f$,
\beq
i\cM =
-4\pi i \al ~\ol{u}(k') \ga^{\mu} u(k) ~\frac{\et_{\mu\nu}}{q^2}
\int d^4x ~e^{iq\cdot x} \langle \X |J^{\nu}(x) |\P\rangle,
\eeq
where $\al$ is the fine structure constant,
$J^{\mu}(x)=\qq f \ol\ps_f(x)\Ga_f^{\mu}\ps_f(x)$
is the quark electromagnetic current
with $\Ga_f^{\mu}=\ga^{\mu}+c^{\nu\mu}_f \ga_{\nu}$,
and where $|\P\rangle$ and $|\X \rangle$ are interaction-picture kets 
for the proton and the final-state hadrons,
respectively.
Squaring this amplitude and summing over all possible final states $X$
gives
\bea
\hskip -20pt 
\sum_\X \int d \Pi_\X |\cM |^2
&
\hskip-5pt 
=
&
\hskip-5pt 
\frac{16 \pi \al^2}{q^4} 
\big[ \ol{u}(k) \ga^{\nu} u(k') ~\ol{u}(k') \ga^{\mu} u(k) \big]
\nonumber\\
&&
\hskip-5pt 
\times 
\sum_\X \int d \Pi_\X \langle \P|J_{\nu}(-q) |\X \rangle 
\langle \X |J_{\mu}(q) |\P\rangle,
\label{eq2}
\eea
where $J^{\mu}(q)$ is the Fourier transform of the current.
After averaging over the spins
as required for an unpolarized cross section,
the term in brackets involving the electron-photon interaction
yields the electron tensor 
$L^{\mu\nu}=
2(k^{\mu}k'^{\nu}+k^{\nu}k'^{\mu}-k\cdot k' \et^{\mu\nu})$. 
The part involving the sum over $X$ 
corresponds to the photon-proton interaction 
and depends on the quark-sector coefficients for Lorentz violation 
via the current $J^{\mu}(q)$. 
It is related to the proton tensor 
\beq
W^{\mu\nu}=
i\int d^4x ~e^{iq\cdot x}\langle \P| J^{\mu}(x)J^{\nu}(0)|\P \rangle
\label{eqW}
\eeq
via the optical theorem,
\beq
\sum_\X \int d \Pi_\X \langle \P|J_{\nu}(-q) |\X \rangle 
\langle \X |J_{\mu}(q) |\P\rangle
= 2 ~\Im W_{\mu\nu},
\eeq
as illustrated in Fig.\ \ref{fig2}.
The proton tensor \rf{eqW} contains 
the information about the photon-proton interaction 
and interactions among the proton constituents. 

To obtain the unpolarized DIS differential cross section 
from the spin-averaged squared amplitude,
we must divide by the flux factor $F$.
Typically,
care is required in defining $F$ in Lorentz-violating situations
because it is frame dependent.
To write $F$ in terms of the Mandelstam variable $s$ 
requires using the momentum-velocity relation, 
which is modified in the presence of Lorentz violation
\cite{ck01}. 
However,
the flux factor here
is associated with the initial electron and proton states,
while the Lorentz violation concerns 
only the internal structure of the proton.
The short-wavelength photon involved in DIS 
sees only the quark substructure,
and so the flux factor can be taken to be 
$F = 2s = 2(k+\P)^2$
as usual.
In effect,
since proton-sector Lorentz violation is well constrained
\cite{tables},
we can neglect it for present purposes.

The differential cross section depends 
on the phase-space variables for the scattered electron.
The trajectory of the final electron lies along a cone
of opening angle $\th$ equal to the scattering angle.
In the Lorentz-invariant case,
the physics is independent of the azimuthal angle $\ph$ around this cone or,
equivalently,
of the orientation of the plane 
defined by the incoming and scattered electron.
However,
in the presence of Lorentz violation,
the physics depends on 
the direction of travel of the scattered quark
relative to $c_f^{\mu\nu}$,
which via energy-momentum conservation 
is equivalent to dependence on $\ph$.
The conventional two dimensionless variables
used to characterize the phase space,
Bjorken $x$ and $y$,
must therefore be supplemented with a third variable
corresponding to $\ph$.
While this third variable might be defined
in a frame-independent way
using products of $c_f^{\mu\nu}$ with the 4-momenta,
it is more transparent for present purposes 
to work directly in a specified frame.
For definiteness,
we adopt a detector frame
having $z$ axis in the direction of the incoming electron,
with the 4-momenta of the incoming electron, 
incoming proton, and outgoing electron 
given by
$k^\mu =  E (1, \hat k)$,
$\P^\mu = E_p ( 1, -\hat k)$,
and 
$k^\prime = E^\prime (1, \hat k^\prime)$,
respectively,
where  
$\hat k = (0, 0, 1)$, 
$\hat k^\prime =( \sin \th \cos \ph, \sin \th \sin\ph,\cos \th)$.
The standard $x$ and $y$ variables are
\bea
x 
&
\hskip-5pt 
=
&
\hskip-5pt 
\frac{-q^2}{2\P\cdot q}
= \frac {E'^2}{4EE_p}\frac{\sin^2\th}{y(1-y)},
\nonumber\\
y
&
\hskip-5pt 
=
&
\hskip-5pt 
\frac{\P\cdot q}{\P\cdot k}
=
1 - \frac {E'}{2E} (1 + \cos\th) ,
\eea
and we choose $x$, $y$, and $\ph$ 
to parametrize the phase space for the differential cross section.

With the above considerations,
the unpolarized differential cross section takes the form 
\beq
\frac{d\si}{dxdyd\ph}=
\frac{\al^2 y}{2\pi q^4}L^{\mu\nu} ~\Im W_{\mu\nu}.
\label{eqCS}
\eeq
The spin average in $W^{\mu\nu}$ is understood.
The form of $W^{\mu\nu}$ can be determined perturbatively, 
as the strong interaction is subdominant for $Q^2=-q^2 \gg M^2$. 
One standard method is to write a general tensorial structure 
for the vectors $\P$ and $q$, 
using the Ward identities to fix its form.
In the presence of Lorentz violation,
this approach is involved 
because the coefficients $c^{\mu\nu}_f$ can also appear. 
Instead,
we proceed below 
using the optical theorem and the Cutkosky rules, 
and we confirm the results via
an alternative approach based on the operator-product expansion (OPE)
in Sec.\ \ref{Operator product expansion}.

\section{Cross section}
\label{sec:results}

The parton model can be used to calculate 
the explicit form of the proton tensor $W^{\mu\nu}$. 
In this approach, 
the photon interacts only with a parton $f$ 
in the state $|\xi \P\rangle$
carrying a fraction $\xi$ of the proton 4-momentum $\P$,
\beq
W^{\mu\nu}\approx 
i \int d^4 x ~e^{iq\cdot x}
\hskip-5pt
\int^1_0 d\xi
\sum_f \frac{f_f(\xi)}{\xi}
\langle \xi \P|J^{\mu}(x)J^{\nu}(0)|\xi \P\rangle.
\label{eq3}
\eeq
Here, 
the parton distribution function (PDF) $f_f(\xi)$ 
is the probability of finding a parton $f$ carrying a momentum $\xi \P$.
Both quarks and antiquarks of species $f$ are included in the sum,
and the notation suppresses the $Q^2$ dependence of the PDF for simplicity. 
The parton propagates according to the Lorentz-violating dispersion relation,
and the integral over $\xi$ allows for all possible fractions of $\P$. 
In this approach,
we take the dominant effects from Lorentz violation as arising
in the matrix element
and neglect possible effects in the PDF.
Studying these effects is an interesting open challenge,
although they are unlikely to change 
the order of magnitudes of the sensitivities estimated below.
Note that the results obtained in this section 
using the parton expression \rf{eq3}
are supported by the OPE evaluation of Eq.\ \rf{eqW}
presented in Sec.\ \ref{Operator product expansion}.

Inserting the explicit form of $J^{\mu}$, 
applying Wick's theorem at zeroth order in $g_s$,
converting to momentum space,
and taking the average over spins
yields
\bea
\hskip-20pt 
W^{\mu\nu}
&
\hskip-5pt 
=
&
\hskip-5pt 
-\half \int^1_0 d\xi \sum_f 
\Q^2_f\fr{f_f(\xi)}{\xi}\fr{\tr[\xi\Ga_f^{\al}\P_{\al}\Ga_f^{\mu}
(\xi\Ga_f^{\be}\P_{\be}+\Ga_f^{\be}q_{\be})\Ga_f^{\nu}]}
{(\xi\Ga_f^{\be}\P_{\be}+\Ga_f^{\be}q_{\be})^2+i\ep}
\nonumber\\
&&
+ (\mu\leftrightarrow \nu, q\leftrightarrow -q ) .
\label{eq5}
\eea
In calculating the trace,
terms beyond first order in $c_f^{\mu\nu}$ can be neglected. 

The propagator denominator can be written as
\bea
(\xi\Ga_f^{\be}\P_{\be}
+\Ga_f^{\be}q_{\be})^2+i\ep
&
\hskip-5pt 
=
&
\hskip-5pt 
(\xi \P+q)^2+2c_f^{\xi \P+q\ \xi \P+q}+i\ep 
\nonumber\\
&
\hskip-5pt 
\approx
&
\hskip-5pt 
-Q^2 + 2\xi \P\cdot q + 2c_f^{qq} 
\nonumber\\ 
&&
+ 2\xi (c_f^{q\P}+c_f^{\P q}) + 2\xi^2 c_f^{\P\P} + i\ep 
\nonumber\\ 
&
\hskip-5pt 
=
&
\hskip-5pt 
\wt{q}^2+2\xi \wt{\P}\cdot \wt{q} + 2\xi^2 c_f^{\P\P}+i \ep,
\eea
where we introduced the notation 
$c_f^{\mu\al} p_{\al}\equiv c_f^{\mu p}$ 
and $\wt{p}^{\mu}\equiv  p^{\mu}+c_f^{\mu p}$. 
We neglect terms proportional to the proton mass $M$
because $M^2\ll Q^2$ and $M^2\ll 2\P\cdot q$ in the DIS limit. 
Note that the term $c_f^{\P\P}$ can be neglected 
only in the rest frame of the proton. 

The sole contribution to the imaginary part of the proton tensor 
comes from the propagator term shown explicitly in Eq.\ \rf{eq5},
\bea
\hskip -20pt
\Im \left( \frac{-1}{(\xi\Ga_f^{\be} \P_{\be}
+\Ga_f^{\be}q_{\be})^2+i\ep}\right) 
&
\hskip-5pt 
=
&
\hskip-5pt 
\pi \de(-\wt{Q}^2+2\xi \wt{\P}\cdot \wt{q} + 2\xi^2 c_f^{\P\P})
\nonumber\\
&&
\hskip-5pt 
=\de_f \de(\xi-x_f'),
\label{de}
\eea
where only the relevant root for $\xi$ and
terms linear in $c_f^{\mu\nu}$ are kept.
Here,
\beq
\de_f= \frac{\pi}{ys}
\left(1-\frac{2}{ys}(c^{\P q}_f+c^{q\P}_f+2xc^{\P\P}_f) \right),
\eeq
and $x_f'=x-x_{f}$ is the Bjorken $x$ 
shifted by a factor 
\beq
x_{f}=\frac 2{ys}(c_f^{qq}+ x c_f^{\P q}+x c_f^{q\P}+ x^2 c_f^{\P\P}),
\eeq
with $s \approx 2k\cdot \P$.
The corresponding delta function for the other propagator term 
has no zeros and so gives no contribution to the DIS cross section. 

To obtain the DIS differential cross section, 
we first contract the imaginary part of the proton tensor 
with $L^{\mu\nu}$
and then evaluate the $\xi$ integral.
This gives
\bea
\hskip -20pt
\frac{d\si}{dxdyd\ph}
&
\hskip-5pt 
=
&
\hskip-5pt 
\frac{\al^2}{q^4} 
\sum_f 
F_{2f}
\Bigg[
\frac {y s^2}{\pi} (1+(1-y)^2)\de_f
+ \frac {y^2 s}{x} x_f 
\nonumber\\
&&
- \frac{4M^2}{s}(c_f^{kk'}+c_f^{k'k})
+ 4 (c_f^{k'\P}+c_f^{\P k'})
+ \frac{4}{x} (1-y) c_f^{kk} 
\nonumber\\
&&
- 4xy c_f^{\P\P} 
-\frac{4}{x} c_f^{k'k'} 
+ 4 (1-y) (c_f^{k\P}+c_f^{\P k}) 
\Bigg],
\label{eq6}
\eea
where 
\beq
F_{2f} = \Q^2_f f_f(x'_f) x'_f
\eeq
represents the contribution from a parton $f$ 
to the proton form factor,
incorporating Lorentz-violation effects. 

For large momentum transfer $Q^2 \gsim m^2_Z$,
where $m_Z$ is the $Z^0$-boson mass,
$Z^0$ exchange must be included in the DIS process of Fig.\ \ref{fig1}.
The squared amplitude $|\cM|^2$ in the expression \rf{eq2}
then takes the form
$|\cM|^2=
|\cM_{\ga}|^2
+\cM^*_{\ga}\cM_{Z}
+\cM^*_{Z}\cM_{\ga}
+|\cM_{Z}|^2$, 
where the indices $\ga$ and $Z^0$ 
denote photon and $Z^0$ exchanges, 
respectively. 
This yields three additional pieces 
for both the proton and the electron tensors.
The propagator $1/q^4$ in the cross section \rf{eqCS} 
is replaced by $(q^2-m_Z^2)/(q^2[(q^2-m_Z^2)^2+\Ga_Z^2 m_Z^2])$
for the two interference terms,
while 
the propagator becomes $1/[(q^2-m_Z^2)^2+\Ga_Z^2 m_Z^2]$
for the $|\cM_{Z}|^2$ term,
where $\Ga_Z$ is the $Z^0$ width. 
After some calculation,
we find the full cross section takes the form 
\bea
\hskip-20pt 
\frac{d\si}{dx dy d\ph}
&
\hskip-5pt 
=
&
\hskip-5pt 
-
\frac{\al^2 y}{\pi q^4} 
\sum_f \frac{\Q^2_f f_f(x'_f)}{x'_f}
\de_f 
L^{\mu\nu}\om^f_{\mu\nu}
\nonumber\\ 
&& 
\hskip-20pt 
-
\frac{\al^2 y(1-m_Z^2/q^2)}{\pi [(q^2-m_Z^2)^2
+\Ga_Z^2 m_Z^2]} 
\sum_f 
\frac{\qq f f_f(x'_f)T^f_3}{x'_f}
\de_f 
\nonumber\\ 
&& 
\times
\left( v_f(\th_W )v_e(\th_W )
L^{\mu\nu}\om^f_{\mu\nu}
-\frac{x'_f s Q^2}{4s^2_W c^2_W}(y-2) \right)
\nonumber\\
&&
\hskip-20pt 
- 
\frac{\al^2 y}{16 \pi[(q^2-m_Z^2)^2
+\Ga_Z^2 m_Z^2]}
\sum_f\frac{f_f(x'_f)}{x'_f}
\de_f
\nonumber\\
&& 
\times
\Bigg[\left( v^2_f(\th_W )+\frac{1}{4s^2_W c^2_W}\right)
\left(v^2_e(\th_W )
+\frac{1}{4s^2_W c^2_W}\right)L^{\mu\nu}\om^f_{\mu\nu} 
\nonumber\\
&&
\hskip15pt 
+\frac{x'_f sQ^2}{s^2_W c^2_W}v_f(\th_W )v_e(\th_W )(y-2)
\Bigg],
\label{xsec}
\eea
where
\bea
\om^{\mu\nu}_f 
&
\hskip-5pt 
=
&
\hskip-5pt 
(\et^{\mu\nu}
+2c_f^{\mu\nu})({x'}_f^{2} \P^2+x'_f \P\cdot q)
-2{x'}_f^{2} \P^{\mu}\P^{\nu}
\nonumber\\
&&
\hskip-5pt 
-[x'_f(2x'_f c_f^{\mu \P}+2x'_f c_f^{\P\mu}+c_f^{q\mu}+c_f^{\mu q})\P^{\nu}
+(\mu \leftrightarrow \nu)] 
\nonumber\\
&&
\hskip-5pt 
+\et^{\mu\nu}(2{x'}_f^{2} c_f^{\P\P}+x'_f c_f^{\P q}+x'_f c_f^{q\P})
+ \ldots, 
\eea
with the ellipsis representing terms 
that vanish when contracted with the electron tensor.
In Eq.\ \rf{xsec},
we abbreviate 
$s_W \equiv \sin \th_W$,
$c_W \equiv \cos \th_W$
and define
\bea
v_u(\th_W)
&
\hskip-5pt 
=
&
\hskip-5pt 
\frac{4\qq u s^2_W-1}{2s_W c_W},
\quad
v_d(\th_W)=-\frac{4\qq d s^2_W+1}{2s_W c_W},
\nonumber\\
v_e(\th_W)
&
\hskip-5pt 
=
&
\hskip-5pt 
\frac{4\qq e s^2_W+1}{2s_W c_W},
\eea
where 
$\qq u= {2}/{3}$, 
$\qq d=-{1}/{3}$, 
$T^u_3={1}/{2}$, 
$T^d_3=-{1}/{2}$.

\section{Operator-product expansion}
\label{Operator product expansion}

To validate our result, 
we can recalculate the cross section using the OPE
to evaluate the product of the two electromagnetic currents 
in the proton tensor \rf{eqW}.
Next, 
this calculation is briefly outlined,
following the presentation of Ref.\ \cite{ps}.

At zeroth order in $g_s$ and for a given flavor $f$,
the dominant terms in the OPE of the two currents take the form
\bea
\hskip-20pt 
\ol{\ps}_f(x)\Ga_f^{\mu}\ps_f(x)~\ol{\ps}_f(0)\Ga_f^{\nu}\ps_f(0)
&
\hskip-5pt 
\supset
&
\hskip-5pt 
\ol{\ps}_f(x)\Ga_f^{\mu} \begC1{\ps_f(x)}
\endC1{~\ol{\ps}_f(0)} \Ga_f^{\nu} \ps_f(0)
\nonumber\\[-5pt] 
&&
\hskip -50pt
+\begC1{\ol{\ps}_f(x)}\conC{\Ga_f^{\mu}\ps_f(x)
~\ol{\ps}_f(0)\Ga_f^{\nu}} \endC1{\ps_f(0)}
+\ldots,
\label{eq10}
\eea
where the field contractions are propagators  
that are singular for $x\rightarrow0$. 
Other possible terms are subdominant at short distances, 
including those involving quark fields of different flavors. 

Implementing the Fourier transform in Eq.\ \rf{eqW}
on the first contraction in Eq.\ \rf{eq10} yields
\vskip-20pt
\bea
\hskip-20pt
\int d^4x ~e^{i q\cdot x} 
\ol{\ps}_f(x)\Ga_f^{\mu}\begC1{\ps_f(x)}\endC1
{~\ol{\ps}_f(0)}\Ga_f^{\nu}\ps_f(0) =
&&
\nonumber\\[-5pt]
&&
\hskip-70pt
\ol{\ps}_f(x)\Ga_f^{\mu} \frac{i(i\wt{\slashed{\prt}}
+\wt{\slashed{q})}}{(i\wt{\prt}+\wt{q})^2}\Ga_f^{\nu}\ps_f(0),
\label{eq11}
\eea
where $q$ is the photon momentum 
and the derivative $i\prt_\mu$ represents 
the momentum carried by the quark field. 
Averaging over the spins symmetrizes this expression in $\mu$ and $\nu$.
In the high-energy limit,
we can write
\beq
\frac{1}{(i\wt{\prt}+\wt{q})^2} 
\approx
-\frac{1}{\wt{Q}^2}
\sum^{\infty}_{n=0}
\left(\frac{2i\wt{q}\cdot\wt{\prt}}{\wt{Q}^2}
\right)^n,
\label{eq12}
\eeq
where we disregard terms involving ${\wt{\prt}^2}/{\wt{Q}^2}$ 
because this is proportional to 
$(\xi\wt{\P})^2/Q^2 = (\xi M)^2/Q^2$ 
and so is negligible for $Q^2\gg M^2$, $2\P\cdot q \gg M^2$. 
The Fourier transform of the second contraction in Eq.\ \rf{eq10} 
involves instead a propagator from $x$ to $0$,
which produces a result of the form \rf{eq12} but with 
the opposite sign for $\wt{q}$.
Combining the two contractions therefore
eliminates all the terms with odd $n$ in the sum \rf{eq12}. 

The proton tensor $W^{\mu\nu}$ is the expectation value of this result,
which requires evaluating products of the momenta $i\wt{\prt}^{\mu}$
in the state $|\P\rangle$.
The dominant contributions to $W^{\mu\nu}$ 
arise from operators of smallest nontrivial twist.
These involve symmetrized indices,
generating a symmetric product of proton momenta $\wt{\P}^{\mu}$,
\beq
\langle \P|\ol{\ps}_f(x)\ga^{(\mu_1}(i\wt{\prt}^{\mu_2})
\ldots (i\wt{\prt}^{\mu_n)})\ps_f(0)|\P\rangle
\approx 2A^n_f \wt{\P}^{\mu_1}\ldots \wt{\P}^{\mu_n}, 
\label{twist}
\eeq 
where $A^n_f$ are the moments of the PDF.
In effect,
this approximation treats the dominant effects of Lorentz violation 
as arising from the replacement 
$\et^{\mu\nu}\to \et^{\mu\nu}+c_f^{\mu\nu}$
in the conventional Lorentz-invariant result.
The moments are observer scalars and unchanged by Lorentz violation,
and the running of the PDF is unaffected by Lorentz violation at this order
\cite{running},
so the PDF maintain standard properties.
We neglect contributions from trace terms for simplicity.
The expression \rf{twist} is gauge invariant
because the derivative $i\wt{\prt}^{\mu}$ is indistinguishable
from the covariant derivative at zeroth order in $\al_s$. 
Dropping terms proportional to $q^{\mu}$
that vanish when contracted with the lepton tensor,
$q^{\mu}L_{\mu\nu}=0$, 
we find the proton tensor takes the form 
\bea
\hskip -20pt
W^{\mu\nu}
&
\hskip-5pt 
\approx
\hskip-5pt 
&
\sum_f
\left[
\P^{\mu}\P^{\nu}+\P^{\mu}(c_f^{\P\nu}+c_f^{\nu \P})
+\P^{\nu}(c_f^{\P\mu}+c_f^{\mu \P})
\right]
W_{2f}
\nonumber\\ 
&&
\hskip-30pt
-\sum_f
\left[
\et^{\mu\nu}
+(c_f^{\mu\nu}+c_f^{\nu\mu})
-\left(
\frac{\P^{\mu }}{\wt{q}\cdot\wt{\P}} (c_f^{\nu q}+c_f^{q\nu})
+(\mu\leftrightarrow\nu) \right)
\right]
W_{1f}.
\nonumber\\ 
\label{WOPE}
\eea
Here,
the partial proton form factors are defined as
\bea
W_{1f}
&
\hskip-5pt 
=
&
\hskip-5pt 
2 \Q^2_f \sum^{\infty}_{n=2}
\frac{(2\wt{q}\cdot\wt{\P})^{n}} {(\wt{Q}^2)^{n}}A^n_f, 
\nonumber\\
W_{2f}
&
\hskip-5pt 
=
&
\hskip-5pt 
8 \Q^2_f \sum^{\infty}_{n=2}
\frac{(2\wt{q}\cdot\wt{\P})^{n-2}}{(\wt{Q}^2)^{n-1}}A^n_f
\label{eqsf}
\eea
with $n$ even.
Modulo trace terms,
they obey a Lorentz-violating version of the Callan-Gross relation
\cite{cg68},
\beq
W_{1f}=\frac{\wt{p}\cdot\wt{q}}{2 x_f'} W_{2f}.
\eeq 
Calculating the differential cross section using Eq.\ \rf{WOPE} 
and identifying 
$\Im W_{2f} = 2\pi F_{2f}/\wt{p}\cdot\wt{q}$
yields 
the result \rf{eq6} obtained via the optical theorem
up to neglected trace terms,
as expected.

\section{Estimated constraints}
\label{sec:limits}

An impressive set of DIS data
for electron-proton and positron-proton scattering
has been accumulated by the H1 and ZEUS collaborations at HERA
\cite{h1zeus}.
These data span six orders of magnitude in both $x$ and $Q^2$.
Here,
we explore the prospects for extracting measurements
of the coefficients $c_f^{\mu\nu}$
from the combined neutral-current cross-section measurements,
involving 644 different values of $x$ and $Q^2$.
Note that these data include measurements at $x\ll 1$
for which some Lorentz-violating effects are enhanced,
as can be seen by inspection of the results \rf{eq6} and \rf{xsec}.
The data were taken at an electron-beam energy of $E = 27.5$ GeV,
while the proton-beam energies used were 
$E_p = 920$ GeV, 820 GeV, 575 GeV, and 460 GeV.

The explicit form of the coefficients $c_f^{\mu\nu}$
for Lorentz violation is frame dependent,
so a standard frame must be adopted to establish results.
The canonical choice  
is the Sun-centered celestial-equatorial frame 
\cite{sunframe},
with cartesian coordinates denoted by $(T,X,Y,Z)$.
The origin for $T$ is specified as the 2000 vernal equinox,
when the $X$ axis points from the Earth to the Sun.
The $Z$ axis matches the rotation axis of the Earth.
The Sun-centered frame is effectively inertial over the time scale
of most laboratory experiments,
including the 15-year period during which the HERA data were obtained.
At HERA,
which is located at longitude $\la\simeq 9.88^\circ$,
the local sidereal time $T_\oplus$ differs from $T$ 
by an offset $T_0 \equiv T-T_\oplus\simeq 3.75$ h
\cite{offset}.
Since the coefficients $c_f^{\mu\nu}$ can be taken symmetric and traceless,
the goal of an experimental analysis is to measure
for each quark flavor $f$
the nine independent components
$c_f^{TX}$,
$c_f^{TY}$,
$c_f^{TZ}$,
$c_f^{XX}$,
$c_f^{XY}$,
$c_f^{XZ}$,
$c_f^{YY}$,
$c_f^{YZ}$,
$c_f^{ZZ}$
expressed in the Sun-centered frame.

The coefficients $c_f^{\mu\nu}$ can be taken as constants
in the Sun-centered frame
\cite{ck}.
The rotation of the Earth then implies that 
in a detector frame they vary with $T_\oplus$ 
at harmonics of the Earth's sidereal frequency 
$\om_\oplus\simeq 2\pi/(23{\rm ~h} ~56{\rm ~min})$
\cite{ak98}.
The differential cross section \rf{xsec},
which is valid in any given detector frame,
therefore also varies with time.
To obtain the explicit form of this time dependence,
we can transform from the Sun-centered frame
to the detector frame.
The transformation can be taken as nonrelativistic 
to an excellent approximation.
At H1, 
the electron beam travels approximately $\vp \simeq 20^\circ$ north of east,
while at ZEUS it travels approximately 20$^\circ$ south of west, 
so the detector frame and therefore the relevant transformation differs
in each case.
The two required transformations are rotations, 
\bea
&
\hskip-30pt 
\cR
=
&
\hskip-15pt 
\left(\begin{array}{rrr}
\pm 1&0 &0 \\
    0 &0 &1 \\
    0 &\mp 1&0 \\
\end{array}\right)
\left(\begin{array}{rrr}
    \cos\vp & \sin \vp &0 \\
    -\sin \vp &\cos \vp &0 \\
    0 & 0 &1 \\
\end{array}\right)
\nonumber\\
&&
\hskip-10pt
\times\left(\begin{array}{rrr}
\cos \chi \cos \om_\oplus T_\oplus
&
\cos \chi \sin \om_\oplus T_\oplus
& 
-\sin \chi 
\\
-\sin \om_\oplus T_\oplus
&
\cos \om_\oplus T_\oplus
& 
0 
\\
\sin \chi \cos \om_\oplus T_\oplus
&
\sin \chi \sin \om_\oplus T_\oplus
& 
\cos \chi
\\
\end{array}\right),
\label{rot}
\eea
where the top and bottom signs hold for H1 and ZEUS, 
respectively,
and where the HERA colatitude is $\ch\simeq 36.4^\circ$.

Under this rotation,
the differential cross section \rf{xsec}
acquires time dependence involving up to second harmonics in $\om_\oplus$.
We can therefore write the integrated cross section as
\bea
\hskip-20pt 
\si (T_\oplus,x,Q^2) 
&
\hskip-5pt 
=
&
\hskip-5pt 
\si_{\rm SM} (x, Q^2) 
\Big(
1 +
c^{\mu\nu}_f \; \al_{\mu\nu}^f   
\nonumber \\
&&
\hskip-5pt 
+ 
c^{\mu\nu}_f \; \be_{\mu\nu}^f  \cos \om_\oplus T_\oplus 
+ c^{\mu\nu}_f  \;\ga_{\mu\nu}^f  \sin \om_\oplus T_\oplus
\nonumber \\
&&
\hskip-5pt 
+ 
c^{\mu\nu}_f  \;\de_{\mu\nu}^f  \cos 2\om_\oplus T_\oplus 
+ c^{\mu\nu}_f \; \ep_{\mu\nu}^f \sin 2\om_\oplus T_\oplus 
\Big),
\label{xs}
\eea
where 
$\al_{\mu\nu}^f$,
$\be_{\mu\nu}^f$,
$\ga_{\mu\nu}^f$,
$\de_{\mu\nu}^f$,
$\ep_{\mu\nu}^f$
are functions of $x$ and $Q^2$
and a sum over $f$ is understood.
These functions can be extracted from the result \rf{xsec}
and are determined in terms of SM parameters 
and the proton PDF.
In what follows,
we adopt values for SM parameters taken from the Particle Data Group
\cite{pdg}
and obtain the proton PDF
using the program ManeParse 
\cite{go15,cl16}
and central values taken from the CT10 set
\cite{la10}.
Many components of the functions 
$\al_{\mu\nu}^f$,
$\be_{\mu\nu}^f$,
$\ga_{\mu\nu}^f$,
$\de_{\mu\nu}^f$,
$\ep_{\mu\nu}^f$
are zero.
For example,
direct consideration of the rotation properties 
associated with the coefficients $c_f^{\mu\nu}$ 
reveals that the combination
$c_f^{TT} \equiv c_f^{XX}+c_f^{YY}+c_f^{ZZ}$
and the components  $c_f^{TZ}$ and $c_f^{ZZ}$
are the only ones that yield
a time-independent contribution to the cross section,
so the functions $\al^f_{\mu\nu}$ are nonvanishing
only for these components.
Similarly,
the cross-section oscillations at frequency $\om_\oplus$
depend only on the components
$c_f^{TX}$, $c_f^{TY}$, $c_f^{YZ}$, and $c_f^{XZ}$,
while the oscillations at $2\om_\oplus$
involves only the component $c_f^{XY}$ 
and the combination $c_f^{XX}-c_f^{YY}$.

Since the HERA data were taken over many years,
any studies of the time-averaged cross section
$\si(x,Q^2) \propto \int dT_\oplus ~\si(T_\oplus,x,Q^2)$
effectively average away 
the possible Lorentz-violating effects 
from the oscillatory terms in the cross section \rf{xs}.
Such studies are thus sensitive only 
to the intrinsically time-independent piece, 
which involves an $x$- and $Q^2$-dependent rescaling 
of the conventional SM cross section $\si_{\rm SM}(x,Q^2)$.
However,
the HERA measurements are included in the global fits 
used to extract the proton PDF,
so the SM predictions for these observables 
necessarily agree with the data.
The absence of an SM prediction independent of the HERA dataset
makes it challenging to extract reliable bounds 
on SME coefficients from the time-averaged cross section,
even taking advantage of the $x$ and $Q^2$ dependence.
In addition,
the components $c^{TT}_u$ and $c^{TT}_u$ 
are related to the ultrarelativistic coefficients
$\ring c_u^{{\rm UR}(4)}$ and $\ring c_d^{{\rm UR}(4)}$
relevant to certain astrophysical studies of Lorentz violation
and are already constrained to exceptional sensitivity, 
$|c_f^{TT}| \equiv |3\ring c_f^{{\rm UR}(4)}/4| \lsim 1.8\times 10^{-21}$
\cite{astro}.  
 
In contrast,
the time-dependent pieces of the cross section \rf{xs}
provide a unique signal for Lorentz violation
involving coefficients that to date
are unconstrained by direct measurement.
A search for sidereal-time dependence of the cross sections
in the HERA dataset
would thereby permit first measurements 
of 12 of the 18 components of the SME coefficients for Lorentz violation
$c_f^{\mu\nu}$, $f=u,d$
expressed in the Sun-centered frame.
In the SM,
when $Q^2$ is large enough for asymptotic freedom to manifest
but small enough so that radiative corrections can be neglected,
the proton structure function exhibits Bjorken scaling,
becoming independent of $Q^2$. 
However, 
the presence of Lorentz violation induces 
a nontrivial {\it tree-level} dependence on $Q^2$ 
in the reduced cross section and the proton structure function.
Here,
we use the sidereal variation and dependence on $x$ and $Q^2$ 
to estimate the sensitivity attainable in a data analysis.

\begin{figure}
\includegraphics[width=\hsize]{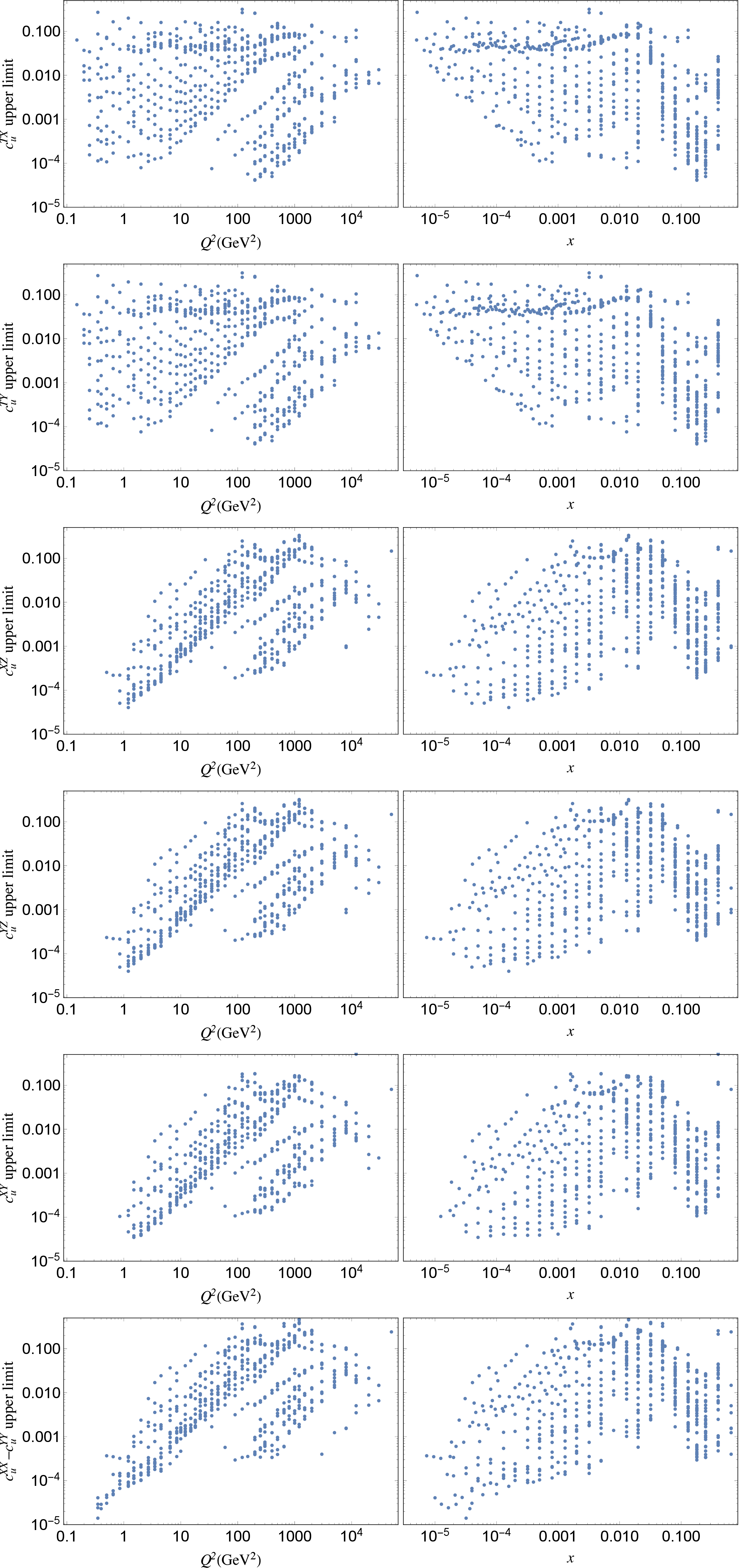}
\caption{Estimated 95\% C.L. constraints on magnitudes
of $c_u^{\mu\nu}$ from each HERA data point.
\label{fig_bounds_up}}
\end{figure}

\begin{figure}
\includegraphics[width=\hsize]{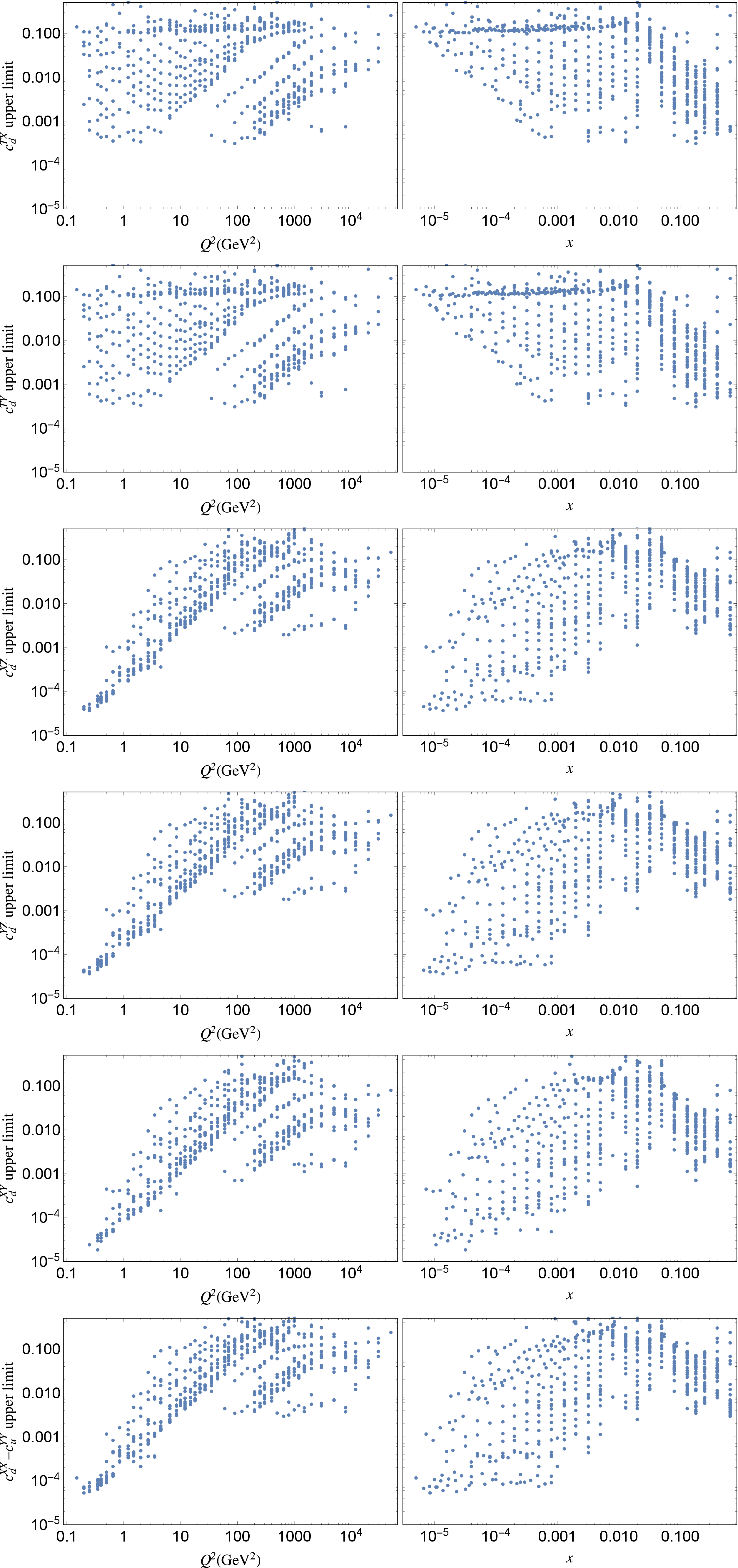}
\caption{Estimated 95\% C.L. constraints on magnitudes
of $c_d^{\mu\nu}$ from each HERA data point.
\label{fig_bounds_down}}
\end{figure}

For simplicity and following accepted procedure
\cite{tables},
we study in turn each of the 12 relevant components of $c_f^{\mu\nu}$,
setting the others to zero.
To explore the potential sensitivity attainable
to studies of sidereal-time variations,
we use the reported HERA value and experimental uncertainty 
for each measured differential cross section
\cite{h1zeus},
as follows.
For each measurement,
we integrate the cross section \rf{xs} 
into four equal-sized bins in sidereal time,
defined starting at $T_\oplus = 0$. 
At fixed $x$ and $Q^2$, 
the integrated cross section
depends on $\si_{\rm SM} (x, Q^2)$ and the chosen component of $c_f^{\mu\nu}$.
For each $x$ and $Q^2$,
we construct a set of 1000 randomized pseudoexperiments,
each with mean cross section equal to the reported value 
and with statistical error per bin 
equal to the reported experimental uncertainty
scaled by the square root of the number of bins.
For each pseudoexperiment,
we construct a $\ch^2$ distribution involving 
four binned measurements and the two fit variables
$\si_{\rm SM} (x, Q^2)$ and $c^{\mu\nu}_f$,
and we extract the 95\% upper bound on 
the chosen component of $|c^{\mu\nu}_f|$.
The estimated constraint on the component
is then taken as the median of the upper bounds over all pseudoexperiments. 

\renewcommand{\arraystretch}{1.1}
\begin{table}
\begin{center}
\setlength{\tabcolsep}{5pt}
\begin{tabular}{ccc}
\hline
\hline
		Coefficient			&				Individual		&				Combined		\\	\hline
$	|	c^{TX}_u	|	$	&	$	<	4	\times 10^{-5}	$	&	$	<	1	\times 10^{-5}	$	\\	
$	|	c^{TY}_u	|	$	&	$	<	4	\times 10^{-5}	$	&	$	<	1	\times 10^{-5}	$	\\	
$	|	c^{XZ}_u	|	$	&	$	<	4	\times 10^{-5}	$	&	$	<	5	\times 10^{-6}	$	\\	
$	|	c^{YZ}_u	|	$	&	$	<	4	\times 10^{-5}	$	&	$	<	5	\times 10^{-6}	$	\\	
$	|	c^{XY}_u	|	$	&	$	<	4	\times 10^{-5}	$	&	$	<	3	\times 10^{-6}	$	\\	
$	|	c^{XX}_u - c^{YY}_u	|	$	&	$	<	1	\times 10^{-5}	$	&	$	<	8	\times 10^{-6}	$	\\	[5pt]
$	|	c^{TX}_d	|	$	&	$	<	3	\times 10^{-4}	$	&	$	<	1	\times 10^{-4}	$	\\	
$	|	c^{TY}_d	|	$	&	$	<	3	\times 10^{-4}	$	&	$	<	1	\times 10^{-4}	$	\\	
$	|	c^{XZ}_d	|	$	&	$	<	4	\times 10^{-5}	$	&	$	<	2	\times 10^{-5}	$	\\	
$	|	c^{YZ}_d	|	$	&	$	<	4	\times 10^{-5}	$	&	$	<	2	\times 10^{-5}	$	\\	
$	|	c^{XY}_d	|	$	&	$	<	2	\times 10^{-5}	$	&	$	<	1	\times 10^{-5}	$	\\	
$	|	c^{XX}_d - c^{YY}_d	|	$	&	$	<	5	\times 10^{-5}	$	&	$	<	3	\times 10^{-5}	$	\\	[3pt]
\hline
\hline
\end{tabular}
\end{center}
\caption{Best estimated 95\% C.L. constraints on magnitudes
of $c_u^{\mu\nu}$ and $c_d^{\mu\nu}$.
\label{tab:results}}
\end{table}

The results of this analysis 
for each of the relevant components of $c^{\mu\nu}_u$ and $c^{\mu\nu}_d$ 
are presented in Figs.\ \ref{fig_bounds_up} and \ref{fig_bounds_down}, 
respectively. 
Each panel shows the expected upper bound 
as a function of $Q^2$ (left) and $x$ (right). 
Each point is constructed 
from one of the 644 neutral-current HERA measurements 
\cite{h1zeus}. 
The strongest individual constraints come mostly 
from measurements at low $x$ and low $Q^2$,
close to the kinematical boundary $Q^2 = s x$,
and they are summarized in Table \ref{tab:results}.
By construction,
the constraints we find are two-sided and symmetric. 
Note that in a real analysis,
more bins could help refine the study of the second harmonics.
Also,
the DIS cross sections for H1 and ZEUS are distinct 
due to the different rotations \rf{rot},
so the two detectors have different sensitivities to Lorentz violation.
However,
as the beam directions at the two detectors are opposite,
keeping only one coefficient component at a time implies 
that the two cross sections are related by a parity transformation,
and they therefore reduce to equivalent forms for the present analysis.

Finally,
we perform a global sidereal-time analysis of the whole HERA dataset,
combining all measurements into a single $\ch^2$ distribution. 
The estimated constraints obtained via this procedure
are also listed in Table \ref{tab:results}. 
They are stronger than the best single-measurement constraints 
because the global analysis takes full advantage 
of correlations between the binned integrated cross sections 
at different values of $x$ and $Q^2$. 

To summarize,
in this work we obtained expressions 
for the Lorentz-violating differential cross section 
for DIS of electrons on protons.
We showed that data taken at HERA
have the potential to place first constraints 
on certain types of Lorentz violation in the quark sector
by searching for oscillations of the measured cross section
at harmonics of the Earth's sidereal frequency.
Our estimates reveal a potential sensitivity 
of parts in a million could be attained
to certain dimensionless quark coefficients $c_f^{\mu\nu}$ 
for Lorentz violation.

Related studies could also be performed
for DIS data for proton-antiproton interactions 
in the Tevatron collider at Fermilab
and for proton-proton interactions in the Large Hadron Collider at CERN.
Incorporating spin dependence in the theory
could also reveal sensitivity to other coefficients
and open the possibility of experimental constraints from polarized DIS,
potentially also including muon-sector effects.
Extending the analysis to include Lorentz violation in the sea 
could allow first constraints on certain gluon coefficients
and other quark flavors. 
The prospects for future direct investigations
of quark-sector Lorentz violation are excellent. 

\section*{Acknowledgments}
This work was supported in part 
by the United States Department of Energy
under grant {DE}-SC0010120,
by the Brazilian Coordena\c c\~ ao de Aperfei\c coamento 
de Pessoal de N\'\i vel Superior
under grant 99999.007290/2015-02,
and by the Indiana University Center for Spacetime Symmetries.

\end{document}